\documentstyle[12pt]{article}
\begin{document}
\baselineskip=15.5pt
\begin{titlepage}

\begin{flushright}
IC/2000/183\\
hep-th/0011290
\end{flushright}
\vspace{10 mm}

\begin{center}
{\Large Closed Universe in Mirage Cosmology}

\vspace{5mm}

\end{center}

\vspace{5 mm}

\begin{center}
{\large Donam Youm\footnote{E-mail: youmd@ictp.trieste.it}}

\vspace{3mm}

ICTP, Strada Costiera 11, 34014 Trieste, Italy

\end{center}

\vspace{1cm}

\begin{center}
{\large Abstract}
\end{center}

\noindent

We study the cosmological evolution of the closed universe on a spherical 
probe brane moving in the AdS$_m\times S^n$ background and the near-horizon 
background of the dilatonic D-branes.   The Friedmann equations describing 
the evolution of the brane universe, and the effective energy density and 
pressure simulated on the probe brane due to its motion in the curved 
background spacetime are obtained and analyzed.  We also comment on the 
relevance of the spherical probe brane to the giant graviton for the 
special value of the probe energy.

\vspace{1cm}
\begin{flushleft}
November, 2000
\end{flushleft}
\end{titlepage}
\newpage

\section{Introduction}

Recently, there has been active investigation on the possibility that 
our four-dimensional universe might be a three-brane embedded in 
higher-dimensional spacetime.  Such idea is motivated by the recent 
proposal \cite{add1,aadd,add2,rs1,rs2,rs3} on solving hierarchy problems 
in particle physics and opens up the possibility of probing the extra 
dimensions in the near future.  According to the brane-world scenario, 
fields of the Standard Model, which are assumed to arise as fluctuations 
of branes in string theories, are confined to the three-brane, while gravity 
can freely propagate in the bulk spacetime.  Even when the extra space is 
infinite, the special property of warped spacetime renders the gravity to 
be localized around the three-brane \cite{rs1,rs2,rs3}.

Lots of work on cosmological models based on the brane-world idea has 
been done, e.g. Refs. \cite{dt,bdl,cgs,cgkt,cf,ftw,kkop}.  Most of models 
assume that the three-brane, in which our four-dimensional universe is 
embedded, evolves with time due to the time evolution of energy density of 
the real matter confined within the three-brane.  In this paper, 
we follow a different approach based on the idea that the cosmological 
evolution of our four-dimensional universe is due to the geodesic 
motion of the (probe) universe three-brane in the curved background of 
other branes in the bulk \cite{cr,cpr,kr,kk}.  In this approach, the motion 
in ambient space induces cosmological evolution of the four-dimensional 
universe on the probe three-brane, simulating various kinds of matter 
responsible for expansion of brane universe \cite{kk}.  One can view such 
simulated matter or `mirage' matter as the contribution of other branes, on 
which hidden gauge interactions are localized, and bulk background fields to 
the cosmological evolution of our four-dimensional universe.

The previous works \cite{kk,pp1,kim1,kim2,pp2,youm} on mirage cosmology 
assume that the probe universe brane is planar, thereby the metric on the 
brane universe becomes that of the expanding flat universe (with zero 
spatial curvature).  In this paper, we consider spherical probe branes 
wrapped around the sphere part of various background spacetimes in string 
theories.  Since the probe universe brane is spherical, the metric on the 
brane universe becomes that of the expanding closed universe (with positive 
spatial curvature).  Just as in the planar probe brane case, the mirage 
energy density consists of various terms simulating massless scalars, 
radiation and superluminal matter (outside of the range of the causality 
condition).  Furthermore, the (total) mirage energy density and pressure 
simulated on the brane universe do not stay always positive during the 
course of cosmological evolution.  One of the consequences of this fact 
is that even if the brane universe is described by the Robertson-Walker 
metric with positive spatial curvature it can expand indefinitely.  Also, 
since the positivity conditions on the energy density and the pressure of 
the singularity theorems of cosmology are violated, one can hope that the 
initial singularity of cosmology may be cured within such framework. 

The spherical probe branes are of interest also because of their relevance 
to the giant gravitons \cite{suss}, proposed to give a bulk interpretation 
of the stringy exclusion principle \cite{exc}.  When the spherical probe 
brane has the same energy as that of a massless particle with the same 
angular momentum and has the stable radius, one can view it as 
gravitons blown up into a sphere.  Therefore, the mirage cosmology 
on a spherical probe brane with constant radius can be regarded as 
an example on a toy cosmological model on a (fuzzy) noncommutative sphere 
\cite{fs}.  We find that the Friedmann equation does not make sense when the 
giant graviton has different radius from the one determined by its angular 
momentum.  This implies that the giant graviton is stable against its radius 
perturbation and it is not possible for a point-like graviton to make 
classical transition to the giant graviton through gradual increase in its 
radius.  

The paper is organized as follows.  In section 2, we discuss the relevant 
aspects of the standard closed expanding universe in arbitrary spacetime 
dimensions.  Then, we study the closed universe mirage cosmology 
in the background of the AdS$_m\times S^n$ spacetime in section 3, and 
in the near-horizon background of the dilatonic D-branes in section 4.

\section{Expanding Closed Universe}

Generally, the induced metric on a spherical probe $p$-brane moving 
in the sphere part of the background spacetime can be put into the 
following standard form for the expanding closed universe in 
$(p+1)$-dimensions:
\begin{equation}
d\tilde{s}^2=-d\eta^2+g(\eta)d\Omega^2_p,
\label{expuvmet}
\end{equation}
where the metric $d\Omega^2_p$ on a unit $p$-sphere is parameterized by 
the angular coordinates $\theta_1,...,\theta_p$ as follows:
\begin{equation}
d\Omega^2_p=d\theta^2_1+\sin^2\theta_1[d\theta^2_2+\sin^2\theta_2(\cdots+
\sin^2\theta_{p-1}d\theta^2_p)].
\label{sphmet}
\end{equation}  
One can further put the metric (\ref{expuvmet}) on the brane universe 
into the following form for the Robertson-Walker metric with positive 
spatial curvature by redefining the coordinate as $\chi=\sin\theta_1$:
\begin{equation}
d\tilde{s}^2=-d\eta^2+g(\eta)\left[{{d\chi^2}\over{1-\chi^2}}+\chi^2
d\Omega^2_{p-1}\right],
\label{rwmet}
\end{equation}
where the metric $d\Omega^2_{p-1}$ is parameterized by $\theta_2,...,
\theta_p$.  So, in this section, we discuss the relevant aspects 
of the standard expanding closed universe in $(p+1)$-dimensions.

To derive the Friedmann equations for the expanding brane universe on a 
spherical probe $p$-brane, we define the scale factor $a$ as $a^2=g$ with 
the Hubble parameter defined as $H=\dot{a}/a$, where the overdot stands for 
the derivative with respect to the cosmic time $\eta$.  We assume that 
the effective energy density $\varrho_{\rm eff}$ and pressure $\wp_{\rm eff}$ 
(as measured in the rest frame), simulated on the probe brane due to its 
motion in the source brane background, are perfect fluids, whose 
energy momentum tensor is generally given by
\begin{equation}
T_{\mu\nu}=(\wp_{\rm eff}+\varrho_{\rm eff})U_{\mu}U_{\nu}+
\wp_{\rm eff}g_{\mu\nu}, 
\label{emtensor}
\end{equation}
where $U_{\mu}$ is the $(p+1)$-velocity of the fluid.  From the Einstein's 
equations ${\cal G}_{\mu\nu}=8\pi T_{\mu\nu}$, where ${\cal G}_{\mu\nu}$ is 
the Einstein tensor of the induced metric (\ref{rwmet}), one obtains the 
following Friedmann equations in the $(p+1)$-dimensional universe on the probe 
brane:
\begin{eqnarray}
\left({{\dot{a}}\over a}\right)^2&=&{{16\pi}\over{p(p-1)}}\varrho_{\rm eff}
-{1\over a^2},
\cr
{{\ddot{a}}\over a}&=&\left[1+{1\over 2}a{\partial\over{\partial a}}\right]
\left({\dot{a}\over a}\right)^2=-{{8\pi}\over {p(p-1)}}((p-2)\varrho_{\rm eff}
+p\wp_{\rm eff}),
\label{frdeqs}
\end{eqnarray}
where the explicit expression for $(\dot{a}/a)^2$ can be found by making 
use of the equations for the probe $p$-brane motion and the defining equation 
for the cosmic time $\eta$.  

From the Friedmann equations (\ref{frdeqs}), one obtains the following 
expressions for the effective matter density and pressure on the brane 
universe:
\begin{eqnarray}
\varrho_{\rm eff}&=&{{p(p-1)}\over{16\pi}}\left[\left({\dot{a}\over a}
\right)^2+{1\over a^2}\right],
\cr
\wp_{\rm eff}&=&-{{p-1}\over{16\pi}}\left[\left(p+a{\partial\over{\partial 
a}}\right)\left({\dot{a}\over a}\right)^2+{{p-2}\over a^2}\right].
\label{effdenprss}
\end{eqnarray}
The scalar curvature of the brane universe is
\begin{eqnarray}
{\cal R}&=&2p\left[{\ddot{a}\over a}+{{p-1}\over 2}
\left({\dot{a}\over a}\right)^2+{{p-1}\over 2}{1\over{a^2}}\right]
=p\left[\left(p+1+a{\partial\over{\partial a}}\right)
\left({\dot{a}\over a}\right)^2+{{p-1}\over a^2}\right]
\cr
&=&4(p-1)\pi(\varrho_{\rm eff}-p\wp_{\rm eff}).
\label{rscsph}
\end{eqnarray}
In particular for the $p=3$ case, the effective pressure and the scalar 
curvature have the following simple expressions in terms of the effective 
matter density:
\begin{eqnarray}
\wp_{\rm eff}&=&-\varrho_{\rm eff}-{1\over 3}a{{\partial\varrho_{\rm eff}}
\over{\partial a}},
\cr
{\cal R}&=&8\pi(4+a\partial_a)\varrho_{\rm eff}.
\label{p3casesph}
\end{eqnarray}

When the effective energy density and pressure satisfy the equation 
of state of the form
\begin{equation}
\wp_{\rm eff}=w\varrho_{\rm eff},
\label{pfeqst}
\end{equation}
where $w$ is a time-independent constant, the conservation of energy 
equation $0=\nabla_{\mu}T^{\mu}_{\ 0}=-\partial_0\varrho_{\rm eff}-
p(\varrho_{\rm eff}+\wp_{\rm eff}){\dot{a}\over a}$ can be integrated 
to give 
\begin{equation}
\varrho_{\rm eff}\propto a^{-p(1+w)}.
\label{rgopta}
\end{equation}
So, the behavior of the effective energy density for particularly 
interesting cases of cosmological fluids are as follows.  For the 
(collisionless, nonrelativistic) dust or matter dominated case, for 
which the pressure is negligible and therefore $w=0$, $\varrho_{\rm eff}
\propto a^{-p}$.  For the radiation or relativistic particles, for 
which $T^{\mu}_{\ \mu}=0$ and therefore $w=1/p$, $\varrho_{\rm eff}
\propto a^{-(p+1)}$.  For the vacuum case, for which $\varrho_{\rm eff}=
-\wp_{\rm eff}$ or $w=-1$, $\varrho_{\rm eff}$ is independent of $a$.  
The causality restricts $w$ to be $|w|\leq 1$.  So, $\varrho_{\rm eff}$ 
satisfying the causality condition can decrease at most as fast as 
$\propto a^{-2p}$ as $a$ increases.  This extreme case is 
characteristic of a massless scalar, for which $\varrho_{\rm eff}=
\wp_{\rm eff}$ or $w=1$.

\section{Brane Cosmology in the AdS$_n\times S^m$ Background}

In this section, we study the mirage cosmology on a spherical 
probe $p$-brane moving in the bulk background of the AdS$_m\times 
S^n$ spacetime.  The radii for AdS$_m$ and $S^m$ are respectively 
$\tilde{L}$ and $L$.  This background spacetime can be obtained from 
string theory or M-theory for the $(p,n)=(3,5),(2,7),(5,4)$ cases with 
$L={{n-3}\over 2}\tilde{L}$ as the near-horizon geometries of the 
D3-, M2- and M5-branes.  Other background spacetime in string theories 
corresponding to the near-horizon geometries of dilatonic D-branes will 
be considered in the next section.  We consider the two separate cases 
of the probe brane wrapping $(i)$ $S^p$ in $S^n$ with $n=p+2$ and $(ii)$ 
$S^p$ in the AdS$_m$ space with $m=p+2$.  
For the former case, we parameterize the metric of the AdS$_m\times S^{p+2}$ 
background spacetime as
\begin{equation}
ds^2={u^2\over\tilde{L}^2}[-dt^2+dx^2_1+\cdots+dx^2_{m-2}]+\tilde{L}^2
{{du^2}\over{u^2}}+L^2d\Omega^2_{p+2},
\label{pp2sphmet}
\end{equation}
where we choose to parameterize the metric $d\Omega^2_{p+2}$ on a unit 
$S^{p+2}$ in the following way:
\begin{equation}
d\Omega^2_{p+2}={1\over{1-\rho^2}}d\rho^2+(1-\rho^2)d\phi^2+\rho^2
d\Omega^2_p,
\label{pp2sphmet2}
\end{equation}
with the metric $d\Omega^2_p$ on a unit $S^p$ parameterized by the 
angular coordinates $\theta_1,...,\theta_p$ as in Eq. (\ref{sphmet}).  
For the cases in which such background spacetime can be obtained as the 
near-horizon geometry of maximally supersymmetric $p$-brane solution of 
the supergravity theories, applying the Hodge-dual transformation when 
necessary, one has the following $(p+1)$-form potential whose magnetic 
field flux threads the $S^{p+2}$ part of the bulk spacetime:
\begin{equation}
A^{p+1}_{\phi\theta_1...\theta_p}=L^{p+1}\rho^{p+1}\sin^{p-1}\theta_1
\cdots\sin\theta_{p-1}\equiv L^{p+1}\rho^{p+1}
\epsilon_{\theta_1...\theta_p},
\label{p1fpsph}
\end{equation}
where $\epsilon_{\theta_1...\theta_p}$ is the volume form of a unit $S^p$.  
For the latter case, we parameterize the metric of the AdS$_{p+2}\times S^n$ 
spacetime as follows:
\begin{equation}
ds^2=-\left(1+{r^2\over\tilde{L}^2}\right)dt^2+{{dr^2}\over
{1+{r^2\over\tilde{L}^2}}}+r^2d\Omega^2_p+L^2d\Omega^2_n, 
\label{adsmet}
\end{equation}
where we parameterize the metric of the unit $S^n$ as $d\Omega^2_n=h_{ij}
(\varphi)d\varphi^id\varphi^j$ ($i,j=1,...,n$) and the metric of the 
unit $S^p$  as in Eq. (\ref{sphmet}).  The $(p+1)$-form potential whose 
electric field flux threading the AdS part of the spacetime is given by
\begin{equation}
A^{p+1}_{t\theta_1...\theta_p}=-{{r^{p+1}}\over\tilde{L}}\sin^{p-1}\theta_1
\cdots\sin\theta_{p-1}\equiv -{r^{p+1}\over\tilde{L}}
\epsilon_{\theta_1...\theta_p}.
\label{p1fpads}
\end{equation}

\subsection{Probe brane in $S^n$}

We consider the spherical probe $p$-brane which wraps $S^p$ in the 
$S^{p+2}$ part of the background AdS$_m\times S^{p+2}$ spacetime.  
Since the AdS$_{p+2}\times S^n$ background generically does not have 
dilaton field, the action for the probe $p$-brane has the form
\begin{equation}
S_p=-T_p\int d^{p+1}\xi\sqrt{-{\rm det}\,\hat{G}_{\alpha\beta}}+
T_p\int d^{p+1}\xi\,\hat{A}^{p+1}.
\label{prbact3}
\end{equation}
In the static gauge, the spatial worldvolume coordinates $\xi^a$ ($a=1,...,p$) 
of the probe $p$-brane are identified with the coordinates of the unit 
$S^p$ in the $S^{p+2}$ part of the source background.  We assume that 
the transverse target space coordinates of the probe $p$-brane depend on the 
time coordinate $\xi^0=t$, only, i.e. the probe brane does not oscillate. 
Due to the translational symmetry of the source brane along the 
$x_i$-directions, the momentum of the probe along these directions is 
conserved, which always enables one go to the frame in which 
$\partial_tx_i=0$.  After Eqs. (\ref{pp2sphmet}) and (\ref{p1fpsph}) are 
substituted and the coordinates of $S^p$ are integrated, the probe action 
takes the following form: 
\begin{eqnarray}
S_p&=&-T_pV_p\int dt\,L^p\rho^p\sqrt{{u^2\over\tilde{L}^2}-{\tilde{L}^2\over 
u^2}(\partial_t u)^2-{L^2\over{1-\rho^2}}(\partial_t \rho)^2-L^2(1-\rho^2)
(\partial_t \phi)^2}
\cr
& &+T_pV_p\int dt\,L^{p+1}\rho^{p+1}\partial_t \phi,
\label{pbactsph}
\end{eqnarray}
where $V_p=2\pi^{{p+1}\over 2}/\Gamma({{p+1}\over 2})$.  

To obtain the equations describing the probe brane motion, we consider the 
following canonical momenta and Hamiltonian of the probe brane:
\begin{eqnarray}
P_u&=&{{\partial{\cal L}}\over{\partial(\partial_t u)}}=
{{T_pV_pL^p\tilde{L}^2\rho^p\partial_t u}\over
{u^2\sqrt{{u^2\over\tilde{L}^2}-{\tilde{L}^2\over u^2}(\partial_t u)^2-
{L^2\over{1-\rho^2}}(\partial_t \rho)^2-L^2(1-\rho^2)(\partial_t\phi)^2}}},
\cr
P_{\rho}&=&{{\partial{\cal L}}\over{\partial(\partial_t \rho)}}=
{{T_pV_pL^{p+2}\rho^p\partial_t \rho}\over{(1-\rho^2)
\sqrt{{u^2\over\tilde{L}^2}-{\tilde{L}^2\over u^2}
(\partial_t u)^2-{L^2\over{1-\rho^2}}(\partial_t \rho)^2-L^2(1-\rho^2)
(\partial_t \phi)^2}}},
\cr
P_{\phi}&=&{{\partial{\cal L}}\over{\partial(\partial_t \phi)}}=
{{T_pV_pL^{p+2}\rho^p(1-\rho^2)\partial_t \phi}\over
{\sqrt{{u^2\over\tilde{L}^2}-{\tilde{L}^2\over u^2}(\partial_t u)^2-
{L^2\over{1-\rho^2}}(\partial_t \rho)^2-L^2(1-\rho^2)(\partial_t \phi)^2}}}
+T_pV_pL^{p+1}\rho^{p+1},
\cr
H&=&E=P_u\partial_t u+P_{\rho}\partial_t \rho+P_{\phi}\partial_t \phi
-{\cal L}
\cr
&=&{{T_pV_pL^p\tilde{L}^{-2}\rho^pu^2}\over{\sqrt{{u^2\over\tilde{L}^2}
-{\tilde{L}^2\over u^2}(\partial_t u)^2-{L^2\over{1-\rho^2}}(\partial_t 
\rho)^2-L^2(1-\rho^2)(\partial_t \phi)^2}}}.
\label{cmhsph}
\end{eqnarray}
We will find in the below that the cosmic scale factor is given by 
$a=L\rho$, which implies that $\partial_t\rho$ has to be nonzero in order 
for the brane universe to evolve with time.  We also assume from now on 
that $\partial_t u=0$, i.e. $u=u_0={\rm const}$.  
Then, making use of the fact that the energy $E$ and the angular momentum 
$P_{\phi}$ of the probe brane is conserved, we obtain the following 
equations describing the motion of the probe brane:
\begin{eqnarray}
(\partial_t\rho)^2&=&{{u^2_0(1-\rho^2)}\over{L^2\tilde{L}^2}}\left[1-
{{(P_{\phi}-T_pV_pL^{p+1}\rho^{p+1})^2+T^2_pV^2_pL^{2(p+1)}\rho^{2p}
(1-\rho^2)}\over{E^2L^2\tilde{L}^2u^{-2}_0(1-\rho^2)}}\right],
\cr
(\partial_t\phi)^2&=&{{(P_{\phi}-T_pV_pL^{p+1}{\rho}^{p+1})^2u^4_0}\over
{E^2L^4\tilde{L}^4(1-\rho^2)^2}}.
\label{eqmtnsph}
\end{eqnarray}

The metric on the brane universe is given by the following induced metric 
on the probe $p$-brane:
\begin{eqnarray}
d\tilde{s}^2&=&-\left[{u^2_0\over\tilde{L}^2}-{L^2\over{1-\rho^2}}
(\partial_t\rho)^2-L^2(1-\rho^2)(\partial_t\phi)^2\right]dt^2
+L^2\rho^2d\Omega^2_p
\cr
&=&-{{T^2_pV^2_pL^{2p}u^4_0}\over{E^2\tilde{L}^4}}\rho^{2p}dt^2+
L^2\rho^2d\Omega^2_p,
\label{indmetsph}
\end{eqnarray}
where Eq. (\ref{eqmtnsph}) is used to simplify the expression.  
In terms of the cosmic time $\eta$ defined through
\begin{equation}
d\eta^2={{T^2_pV^2_pL^{2p}u^4_0}\over{E^2\tilde{L}^4}}\rho^{2p}dt^2,
\label{cstmsph}
\end{equation}
the induced metric takes the standard form (\ref{expuvmet}) or (\ref{rwmet}) 
for the expanding closed universe with $g=a^2=L^2\rho^2$.  

The Friedman equations describing the expanding universe on the probe 
$p$-brane can be obtained by substituting Eqs. (\ref{eqmtnsph}) and 
(\ref{cstmsph}) into the following:
\begin{equation}
\left({\dot{a}\over a}\right)^2={1\over\rho^2}\left({{dt}\over{d\eta}}
\right)^2\left({{d\rho}\over{dt}}\right)^2, \ \ \ \ \ \ \ \ \ 
{{\ddot{a}}\over a}=\left[1+{1\over 2}a{\partial\over{\partial a}}\right]
\left({\dot{a}\over a}\right)^2.
\label{frdeqsphfm}
\end{equation}
The resulting expressions for Friedmann equations in terms of the cosmic
scale factor are
\begin{equation}
\left({\dot{a}\over a}\right)^2
={1\over L^2}\left[(\bar{E}^2-\bar{P}^2_{\phi})\left({L\over a}
\right)^{2(p+1)}-\bar{E}^2\left({L\over a}\right)^{2p}+2\bar{P}_{\phi}
\left({L\over a}\right)^{p+1}-\left({L\over a}\right)^2\right],
\label{frdeqsph}
\end{equation}
\begin{equation}
{\ddot{a}\over a}=-{1\over L^2}\left[p(\bar{E}^2-\bar{P}^2_{\phi})
\left({L\over a}\right)^{2(p+1)}-(p-1)\bar{E}^2\left({L\over a}
\right)^{2p}+(p-1)\bar{P}_{\phi}\left({L\over a}\right)^{p+1}\right],
\label{2frdeqsph}
\end{equation}
where we introduced $N\equiv T_pV_pL^{p+1}$, $\bar{P}_{\phi}\equiv 
P_{\phi}/N$ and $\bar{E}\equiv E\tilde{L}L/(Nu_0)$ to simplify the 
expressions.  

The following effective energy density and pressure on the brane 
universe are obtained by substituting Eq. (\ref{frdeqsph}) into Eq. 
(\ref{effdenprss}):
\begin{eqnarray}
\varrho_{\rm eff}&=&{{p(p-1)}\over{16\pi L^2}}\left[(\bar{E}^2-
\bar{P}^2_{\phi})\left({L\over a}\right)^{2(p+1)}-\bar{E}^2\left({L\over a}
\right)^{2p}+2\bar{P}_{\phi}\left({L\over a}\right)^{p+1}\right],
\cr
\wp_{\rm eff}&=&{{p-1}\over{16\pi L^2}}\left[(p+2)(\bar{E}^2-
\bar{P}^2_{\phi})\left({L\over a}\right)^{2(p+1)}-
p\bar{E}^2\left({L\over a}\right)^{2p}+2\bar{P}_{\phi}\left({L\over a}
\right)^{p+1}\right].
\label{effdnsph}
\end{eqnarray}
The following scalar curvature of the brane universe is obtained by 
substituting Eq. (\ref{frdeqsph}) into Eq. (\ref{rscsph}):
\begin{equation}
{\cal R}={p\over L^2}\left[(p+1)(\bar{P}^2_{\phi}-\bar{E}^2)
\left({L\over a}\right)^{2(p+1)}+(p-1)\bar{E}^2\left({L\over a}\right)^{2p}
\right].
\label{sccuvsph}
\end{equation}

As can be seen in Eq. (\ref{effdnsph}), the effective energy density 
$\varrho_{\rm eff}$ simulated on the probe brane consists of $(i)$ 
acausal term $\sim a^{-2(p+1)}$, $(ii)$ term $\sim a^{-2p}$ simulating 
a massless scalar and $(iii)$ term $\sim a^{-(p+1)}$ simulating 
radiation or relativistic matter.  At very early times of the evolution, 
when the probe brane is very close to the source brane (i.e. $a\ll 1$), 
the acausal term dominates the effective energy density and the brane 
expansion, causing superluminal ``shocks''.  As the brane universe expands, 
the massless scalar term $\sim a^{-2(p+1)}$ takes over, and then finally at 
later stage of the evolution the probe brane motion simulates 
radiation-dominated universe.  Note, since the coordinate $\rho=a/L$ in the 
metric (\ref{pp2sphmet2}) is restricted to take value $0\leq\rho\leq 1$, the 
brane universe cannot expand indefinitely.  As can be seen from Eq. 
(\ref{frdeqsph}), the cosmic scale factor $a$ can never reach $L$ 
(since the RHS of Eq. (\ref{frdeqsph}) becomes negative when $a=L$),  
but reaches some maximum value $a_{\rm max}<L$ and starts decreasing 
(since $\ddot{a}<0$ when $\dot{a}=0$), eventually the brane universe 
crunching to zero size.  The scalar curvature (\ref{sccuvsph}) diverges 
when $a=0$ (initial singularity).  However, from the higher dimensional 
perspective (of the bulk background spacetime), such initial singularity 
corresponds to a perfectly regular point in $S^{p+2}$, which is regular 
everywhere, thereby the initial singularity of the brane universe is 
resolved through higher-dimensional embedding.

It is argued in Ref. \cite{suss} that gravitons moving in the AdS$_m\times 
S^n$ spacetime along the $S^n$ part blow up into spherical $(n-2)$-brane of 
increasing size as their angular momentum increases by showing that the 
spherical $(n-2)$-brane with the stable radius (that does not change with 
time) for a give angular momentum has the same energy as that of graviton 
with the same angular momentum.  So, when the energy $E$ of the probe 
$p$-brane (under consideration in this paper) has the same energy as that 
of graviton with the same angular momentum $P_{\phi}$, one can view the 
probe as the `giant gravitons'.  To determine the criterion for the probe 
brane to become the expanded gravitons, we consider the Hamiltonian in Eq. 
(\ref{cmhsph}) expressed in terms of conjugate momenta in the following way:
\begin{equation}
H={u\over\tilde{L}}\left[{P^2_{\phi}\over L^2}+{P^2_u\over{\tilde{L}_2/u^2}}+
{P^2_{\rho}\over{L^2/(1-\rho^2)}}+{{(\rho P_{\phi}-N\rho^p)^2}\over
{L^2(1-\rho^2)}}\right]^{1\over 2}.
\label{ggham}
\end{equation}
Note, we are interested in the probe motion with $u=u_0={\rm const}$ in 
this subsection, so $P_u=0$ in the above Hamiltonian.  Since we are looking 
for the probe motion with the stable radius $\rho$, $P_{\rho}=0$.  For such 
case, the energy takes the following minimum value
\begin{equation}
H={{u_0P_{\phi}}\over{L\tilde{L}}},
\label{minensph}
\end{equation}
when $\rho$ takes the equilibrium value determined by
\begin{equation}
P_{\phi}=N\rho^{p-1}.
\label{eqradrho}
\end{equation}
The minimum energy (\ref{minensph}) of the probe $p$-brane coincides with 
the energy of a massless particle, namely graviton, carrying the same angular 
momentum $P_{\phi}$ on $S^{p+2}$, thereby the probe brane behaves like a 
massless particle or a graviton for such case.  Since the allowed range of 
the coordinate $\rho$ in the metric (\ref{pp2sphmet2}) is $0\leq\rho\leq 1$, 
the maximum value of angular momentum for such particular probe motion is 
$P_{\phi}=N$, as explained in Ref. \cite{suss} as the bulk origin of the 
stringy exclusion principle \cite{exc}.  We notice that when the energy $E$ 
of the probe takes the minimum value (\ref{minensph}) the acausal terms in 
Eqs. (\ref{frdeqsph}-\ref{sccuvsph}) vanish, since $|\bar{E}|=
|\bar{P}_{\phi}|$ for such case.  The first Friedmann equation 
(\ref{frdeqsph}) then takes the following form:
\begin{equation}
\left({\dot{a}\over a}\right)^2=-{1\over L^2}
\left[\bar{P}_{\phi}\left({L\over a}\right)^p-{L\over a}\right]^2.
\label{eepfrdeqsph}
\end{equation}
The RHS of this equation becomes negative unless $a=L
\bar{P}^{1/(p-1)}_{\phi}$, which corresponds to the radius $\rho_0=
(P_{\phi}/N)^{1/(p-1)}$ of the giant graviton.  From this, one can see that 
the giant graviton is stable against its radius perturbation.  Eq. 
(\ref{eepfrdeqsph}) also implies that classical transition from a point-like 
graviton to the giant graviton cannot occur through gradual increase in the 
size of a graviton.  The transition is possible only through the quantum 
tunneling between two vacua, one corresponding to a point-like graviton and 
the other the giant graviton, described by instantons, as was proposed in 
Refs. \cite{ads1,ads2}.

\subsection{Probe brane in AdS}

We consider the spherical probe $p$-brane which wraps $S^p$ in the 
AdS$_{p+2}$ part of the AdS$_{p+2}\times S^n$ background space.  
The action for the probe brane has the form (\ref{prbact3}).  
After Eqs. (\ref{adsmet}) and (\ref{p1fpads}) are substituted and the  
coordinates of $S^p$ are integrated, the probe action (\ref{prbact3}) 
in the static gauge takes the following form:
\begin{equation}
S_p=-T_pV_p\int dt\,r^p\sqrt{1+{r^2\over\tilde{L}^2}-{{(\partial_t r)^2}
\over{1+{r^2\over\tilde{L}^2}}}-L^2h_{ij}\partial_t\varphi^i
\partial_t\varphi^j}+T_pV_p\int dt\,{r^{p+1}\over\tilde{L}},
\label{intphprbact}
\end{equation}
where the sign of the Wess-Zumino term has been reversed, corresponding 
to choosing the opposite brane charge.  

To obtain the equations describing the probe brane motion, we consider the 
following conjugate momenta and Hamiltonian of the probe $p$-brane:
\begin{eqnarray}
P_r&=&{{\partial{\cal L}}\over{\partial(\partial_t r)}}=
{{T_pV_pr^p\partial_t r}\over{(1+{r^2\over\tilde{L}^2})
\sqrt{1+{r^2\over\tilde{L}^2}-{(\partial_t r)^2\over
{1+{r^2\over\tilde{L}^2}}}-L^2h_{ij}\partial_t \varphi^i
\partial_t\varphi^j}}},
\cr
P_i&=&{{\partial{\cal L}}\over{\partial(\partial_t \varphi^i)}}=
{{T_pV_pL^2r^ph_{ij}\partial_t \varphi^j}\over
{\sqrt{1+{r^2\over\tilde{L}^2}-{{(\partial_t r)^2}\over
{1+{r^2\over\tilde{L}^2}}}-L^2h_{ij}\partial_t \varphi^i
\partial_t \varphi^j}}},
\cr
H&=&E=P_r\partial_t r+P_i\partial_t \varphi^i-{\cal L}
\cr
&=&T_pV_p\left[{{r^p\left(1+{r^2\over\tilde{L}^2}\right)}
\over{\sqrt{1+{r^2\over\tilde{L}^2}-{{(\partial_t r)^2}\over
{1+{r^2\over\tilde{L}^2}}}-L^2h_{ij}\partial_t\varphi^i
\partial_t\varphi^j}}}-{r^{p+1}\over\tilde{L}}\right].
\label{cjmhm}
\end{eqnarray}
Making use of the fact that the energy $E$ and the total angular momentum 
$h^{ij}P_iP_j=\ell^2$ of the probe are conserved, from Eq. (\ref{cjmhm}) one 
obtains the following equation describing the radial motion of the probe:
\begin{equation}
(\partial_t r)^2=\left(1+{r^2\over\tilde{L}^2}\right)^2\left[1-
{{1+{r^2\over\tilde{L}^2}}\over{\left({r^{p+1}\over\tilde{L}}+
{E\over{T_pV_p}}\right)^2}}{{T^2_pV^2_pL^2r^{2p}+\ell^2}\over
{T^2_pV^2_pL^2}}\right],
\label{eqrdmtnads}
\end{equation}
along with
\begin{equation}
h_{ij}\partial_t\varphi^i\partial_t\varphi^j={{\left(1+{r^2\over\tilde{L}^2}
\right)^2\ell^2}\over{T^2_pV^2_pL^4\left({r^{p+1}\over\tilde{L}}
+{E\over{T_pV_p}}\right)^2}}.
\label{angmtnads}
\end{equation}

The metric of the brane universe is given by the following induced metric 
on the probe $p$-brane:
\begin{eqnarray}
d\tilde{s}^2&=&-\left[1+{r^2\over\tilde{L}^2}-{{(\partial_t r)^2}\over{1+
{r^2\over\tilde{L}^2}}}-L^2h_{ij}\partial_t\varphi^i\partial_t\varphi^j
\right]dt^2+r^2d\Omega^2_p
\cr
&=&-{{\left(1+{r^2\over\tilde{L}^2}\right)^2r^{2p}}\over
{\left({r^{p+1}\over\tilde{L}}+{E\over{T_pV_p}}\right)^2}}dt^2
+r^2d\Omega^2_p,
\label{brnmet2}
\end{eqnarray}
where we used Eqs. (\ref{eqrdmtnads}) and (\ref{angmtnads}) to simplify the 
expression.  In terms of the cosmic time $\eta$ defined through
\begin{equation}
d\eta^2={{\left(1+{r^2\over\tilde{L}^2}\right)^2r^{2p}}\over
{\left({r^{p+1}\over\tilde{L}}+{E\over{T_pV_p}}\right)^2}}dt^2,
\label{cstmads}
\end{equation}
the induced metric (\ref{brnmet2}) takes the standard form (\ref{expuvmet}) 
for the metric of the expanding closed universe with $g(\eta)=r^2(\eta)$ 
(therefore, the cosmic scale factor is $a=\sqrt{g}=r$).  
Here, just as in the previous section,  $d\Omega^2_p$ is given by Eq. 
(\ref{sphmet}) and through the redefinition of the coordinate $\chi=\sin\psi$ 
the induced metric can be put into the standard form (\ref{rwmet}) for the 
Robertson-Walker metric.  

The Friedman equations describing the expanding brane universe on the 
spherical probe $p$-brane can be obtained by substituting Eqs. 
(\ref{eqrdmtnads}) and (\ref{cstmads}) into the following:
\begin{equation}
\left({\dot{a}\over a}\right)^2={1\over r^2}\left({{dt}\over{d\eta}}
\right)^2\left({{dr}\over{dt}}\right)^2, \ \ \ \ \ \ \ \ \ 
{{\ddot{a}}\over a}=\left[1+{1\over 2}a{\partial\over{\partial a}}\right]
\left({\dot{a}\over a}\right)^2.
\label{frdeqadsfm}
\end{equation}
The resulting expressions are given by 
\begin{equation}
\left({\dot{a}\over a}\right)^2
={1\over\tilde{L}^2}\left[(\tilde{E}^2-\tilde{\ell}^2)\left({\tilde{L}
\over a}\right)^{2(p+1)}-\tilde{\ell}^2\left({\tilde{L}\over a}\right)^{2p}
+2\tilde{E}\left({\tilde{L}\over a}\right)^{p+1}-\left({\tilde{L}\over 
a}\right)^2\right],
\label{frdeqsads}
\end{equation}
\begin{equation}
{\ddot{a}\over a}=-{1\over\tilde{L}^2}\left[p(\tilde{E}^2-\tilde{\ell}^2)
\left({\tilde{L}\over a}\right)^{2(p+1)}-(p-1)\tilde{\ell}^2
\left({\tilde{L}\over a}\right)^{2p}
+(p-1)\tilde{E}\left({\tilde{L}\over a}\right)^{p+1}\right],
\label{2frdeqsads}
\end{equation}
where we introduced $\tilde{N}\equiv T_pV_pL\tilde{L}^p$, $\tilde{\ell}
\equiv\ell/\tilde{N}$ and $\tilde{E}\equiv EL/\tilde{N}$ to simplify the 
expressions.  These Friedmann equations have the similar forms as those 
(\ref{frdeqsph},\ref{2frdeqsph}) associated with the probe brane moving 
in the $S^n$ part of the bulk background except that the massless scalar 
and the radiation terms are now respectively controlled by the angular 
momentum and the energy.  

By substituting Eq. (\ref{frdeqsads}) into Eq. (\ref{effdenprss}), one 
obtains the following effective energy density and pressure on the probe brane:
\begin{eqnarray}
\varrho_{\rm eff}&=&{{p(p-1)}\over{16\pi\tilde{L}^2}}\left[(\tilde{E}^2-
\tilde{\ell}^2)\left({\tilde{L}\over a}\right)^{2(p+1)}-\tilde{\ell}^2
\left({\tilde{L}\over a}\right)^{2p}
+2\tilde{E}\left({\tilde{L}\over a}\right)^{p+1}\right].
\cr
\wp_{\rm eff}&=&{{p-1}\over{16\pi\tilde{L}^2}}\left[(p+2)(\tilde{E}^2-
\tilde{\ell}^2)\left({\tilde{L}\over a}\right)^{2(p+1)}-
p\tilde{\ell}^2\left({\tilde{L}\over a}\right)^{2p}
+2\tilde{E}\left({\tilde{L}\over a}\right)^{p+1}\right].
\label{denads}
\end{eqnarray}
The scalar curvature of the brane universe is
\begin{equation}
{\cal R}={p\over\tilde{L}^2}\left[(p+1)(\tilde{\ell}^2-\tilde{E}^2)
\left({\tilde{L}\over a}\right)^{2(p+1)}+(p-1)\tilde{\ell}^2
\left({\tilde{L}\over a}\right)^{2p}\right].
\label{curvads}
\end{equation}

The effective energy density in Eq. (\ref{denads}) consists of $(i)$ 
acausal term $\sim a^{-2(p+1)}$, $(ii)$ term $\sim a^{-2p}$ simulating 
a massless scalar and $(iii)$ term $\sim a^{-(p+1)}$ simulating radiation or 
relativistic matter.  Just as in the previous subsection, the brane universe 
expansion is dominated by the acausal term during the very early 
stage of cosmological evolution and as the brane universe expands the 
massless scalar term and then the radiation term takes over.  Unlike the 
case in the previous subsection, the coordinate $r=a$ in the metric 
(\ref{adsmet}) can take arbitrarily large values.  So, the brane universe 
can expand indefinitely even if it is described by the metric 
(\ref{expuvmet}) of the expanding closed universe.  This is because the 
effective pressure $\wp_{\rm eff}$ does not always stay positive in the 
course of cosmic evolution of the brane universe.  When $a=0$, the scalar 
curvature (\ref{curvads}) diverges, corresponding to the initial singularity.  
From the higher-dimensional perspective, this initial singularity corresponds 
to a regular point in the AdS part of the AdS$_{p+2}\times S^n$ background 
spacetime. 

It is argued \cite{ads1,ads2} that gravitons can expand also into the 
spherical part of the AdS portion of the AdS$\times S^n$ spacetime.  
To determine the condition for the spherical probe $p$-brane (moving 
in the AdS portion of the AdS$\times S^n$ spacetime) to become giant 
gravitons, we consider the following Hamiltonian for the probe brane 
in equilibrium radius $r$, i.e. Eq. (\ref{cjmhm}) with $\partial_tr=0$:
\begin{eqnarray}
H&=&T_pV_p\left[{1\over{T_pV_pL}}\sqrt{\left(1+{r^2\over\tilde{L}^2}
\right)\left(\ell^2+T^2_pV^2_pL^2r^{2p}\right)}-{r^{p+1}\over\tilde{L}}
\right]
\cr
&=&{\tilde{N}\over L}\left[\sqrt{\left(1+{r^2\over\tilde{L}^2}\right)
\left(\tilde{\ell}^2+{r^{2p}\over\tilde{L}^{2p}}\right)}
-{r^{p+1}\over\tilde{L}^{p+1}}\right].
\label{cstradhamads}
\end{eqnarray}
The Hamiltonian takes the following stable minimum value
\begin{equation}
H={\ell\over L},
\label{minhamads}
\end{equation}
when $r$ takes the equilibrium value determined by 
\begin{equation}
\left({r\over\tilde{L}}\right)^{p-1}=\tilde{\ell}.
\label{eqlrads}
\end{equation}
The minimum energy (\ref{minhamads}) coincides with the energy of a massless 
particle with the angular momentum $\ell$ moving in the AdS space, thereby 
the spherical probe $p$-brane with the energy (\ref{minhamads}) can be 
identified as gravitons expanded into a $p$-sphere.   On the other hand, 
since the range of radial coordinate $r$ in the metric (\ref{adsmet}) is not 
restricted, the probe angular momentum $\ell$ satisfying (\ref{eqlrads}), 
i.e. $\ell=N(r/\tilde{L})^{p-1}=T_pV_pL\tilde{L}r^{p-1}$, can take any 
arbitrarily large values \cite{ads1,ads2}.  As in the previous subsection, 
the minimum energy (\ref{minhamads}) corresponds to the critical energy for 
which the acausal terms in Eqs. (\ref{frdeqsads}-\ref{curvads}) vanish, 
i.e. $|\tilde{E}|=|\tilde{\ell}|$ and the first Friedmann equation 
(\ref{frdeqsads}) takes the form (\ref{eepfrdeqsph}) with $\bar{P}_{\phi}$ 
and $L$ respectively replaced by $\tilde{\ell}$ and $\tilde{L}$.

\section{Brane Cosmology in Near-Horizon Background of Dilatonic D-brane}

In this section, we study the mirage cosmology on the spherical probe 
D$p$-brane moving in the bulk background of the near-horizon region of the 
source D$(6-p)$-brane, where the probe brane wraps $S^p$ in the transverse 
space of the source brane.  

The solution for the source D$(6-p)$-brane magnetically charged under 
the RR $(p+1)$-form potential $A^{p+1}$ has the form
\begin{eqnarray}
ds^2&=&-g_{tt}dt^2+\sum^{6-p}_{i=1}g_{ii}dx^2_i+g_{rr}dr^2
+h(r)r^2d\Omega^2_{p+2},
\cr
e^{\Phi}&=&\left(\textstyle{L_p\over r}\right)^{{(p-3)(p+1)}\over 4},
\ \ \ \ \ \ 
A^{p+1}_{\phi\theta_1...\theta_p}=L^{p+1}_p\rho^{p+1}
\epsilon_{\theta_1...\theta_p},
\cr
g_{tt}&=&\left(\textstyle{r\over L_p}\right)^{{p+1}\over 2},\ \ 
g_{ii}=\left(\textstyle{r\over L_p}\right)^{{p+1}\over 2},\ \ 
g_{rr}=\left(\textstyle{L_p\over r}\right)^{{p+1}\over 2},\ \ 
h(r)=\left(\textstyle{L_p\over r}\right)^{{p+1}\over 2},
\label{nrdbrnmet}
\end{eqnarray}
where $d\Omega^2_{p+2}$ is parameterized as in Eq. (\ref{pp2sphmet2}).  
$L_p$ can be expressed in terms of the number $N$ of D$(6-p)$-branes and 
the tension $T_p$ of the probe D$p$-brane as $L^{p+1}_p=N/(T_pV_p)$.   

The action for the probe D$p$-brane in the absence of the NS $B$-field and 
the worldvolume $U(1)$ gauge field is given by
\begin{equation}
S_p=-T_p\int d^{p+1}\xi\,e^{-\Phi}\sqrt{-{\rm det}\,\hat{G}_{\alpha\beta}}+
T_p\int d^{p+1}\xi\,\hat{A}^{p+1},
\label{prbdpbrnact}
\end{equation}
where $\hat{G}_{\alpha\beta}$ and $\hat{A}^{p+1}$ are respectively the 
pullbacks of the metric and the RR $(p+1)$-form potential.  In the static 
gauge, the spatial worldvolume coordinates $\xi^a$ ($a=1,...,p$) of the 
probe D$p$-brane are identified with the coordinates of the unit $S^p$ 
in the transverse space of the source D$(6-p)$-brane.  As in the previous 
section, we assume that the transverse target space coordinates of the probe 
$p$-brane depend on the time coordinate $\xi^0=t$, only.  We consider the 
frame in which $\partial_tx_i=0$.  
After Eq. (\ref{nrdbrnmet}) is substituted and the coordinates of $S^p$ 
are integrated, the probe action (\ref{prbdpbrnact}) takes the following form:
\begin{equation}
S=-T_pV_p\int dt\,e^{-\Phi}[h(r)r^2\rho^2]^{p\over 2}
\sqrt{g_{tt}-g_{rr}(\partial_t r)^2-g_{\rho\rho}(\partial_t\rho)^2
-g_{\phi\phi}(\partial_t\phi)^2}+N\int dt\,\rho^{p+1}\partial_t\phi.
\label{prbact2}
\end{equation}

The conjugate momenta and the Hamiltonian of the probe brane are given by
\begin{eqnarray}
P_r&=&{{\partial{\cal L}}\over{\partial(\partial_t r)}}=
{{T_pV_pe^{-\Phi}}\over\sqrt{g_{tt}-g_{rr}(\partial_t r)^2-g_{\rho\rho}
(\partial_t \rho)^2-g_{\phi\phi}(\partial_t\phi)^2}}(hr^2\rho^2)^{p\over 2}
g_{rr}\partial_t r,
\cr
P_{\rho}&=&{{\partial{\cal L}}\over{\partial(\partial_t \rho)}}=
{{T_pV_pe^{-\Phi}}\over\sqrt{g_{tt}-g_{rr}(\partial_t r)^2
-g_{\rho\rho}(\partial_t \rho)^2-g_{\phi\phi}(\partial_t \phi)^2}}
(hr^2\rho^2)^{p\over 2}g_{\rho\rho}\partial_t\rho,
\cr
P_{\phi}&=&{{\partial{\cal L}}\over{\partial(\partial_t\phi)}}=
{{T_pV_pe^{-\Phi}}\over\sqrt{g_{tt}-g_{rr}(\partial_tr)^2-g_{\rho\rho}
(\partial_t\rho)^2-g_{\phi\phi}(\partial_t\phi)^2}}(hr^2\rho^2)^{p\over 2}
g_{\phi\phi}\partial_t\phi+N\rho^{p+1},
\cr
H&=&E=P_r\partial_tr+P_{\rho}\partial_t\rho+P_{\phi}\partial_t\phi-{\cal L}
\cr
&=&{{T_pV_pe^{-\Phi}(hr^2\rho^2)^{p\over 2}g_{tt}}\over
{\sqrt{g_{tt}-g_{rr}(\partial_tr)^2-g_{\rho\rho}(\partial_t\rho)^2
-g_{\phi\phi}(\partial_t\phi)^2}}}.
\label{canmomhampp6}
\end{eqnarray}
The metric of the brane universe is given by the following induced 
metric on the probe D$p$-brane:
\begin{equation}
d\tilde{s}^2=-\left[g_{tt}-g_{rr}(\partial_tr)^2-g_{\rho\rho}
(\partial_t\rho)^2-g_{\phi\phi}(\partial_t\phi)^2\right]dt^2
+h(r)r^2\rho^2 d\Omega^2_p.
\label{idmetpm6}
\end{equation}
By defining the cosmic time $\eta$ through
\begin{equation}
d\eta^2=\left[g_{tt}-g_{rr}(\partial_tr)^2-g_{\rho\rho}(\partial_t\rho)^2
-g_{\phi\phi}(\partial_t\phi)^2\right]dt^2,
\label{costime}
\end{equation}
one can put the induced metric (\ref{idmetpm6}) into the standard form 
(\ref{expuvmet}) or (\ref{rwmet}) for the expanding closed universe with 
$g(\eta)=h(r(\eta))r^2(\eta)\rho^2(\eta)$.  Since the scale factor is 
given by $a=\sqrt{g}=h^{1/2}(r)r\rho$, the Friedmann equations 
are given by
\begin{equation}
\left({{\dot{a}}\over a}\right)^2=\left[{{\dot{r}}\over r}\left({1\over 2}
{{h^{\prime}}\over h}r+1\right)+{\dot{\rho}\over\rho}\right]^2,
\ \ \ \ \ \ \ \ \ 
{{\ddot{a}}\over a}=\left[1+{1\over 2}a{\partial\over{\partial a}}\right]
\left({\dot{a}\over a}\right)^2,
\label{lbzrslt}
\end{equation}
where the prime stands for derivative with respect to $r$.  
In studying the Friedman equations, in the following we consider the two 
particular cases of the probe motion with either $\dot{r}=0$ or 
$\dot{\rho}=0$. 

First, when $\partial_tr=0$, i.e. $r=r_{0}={\rm const}$, from Eq. 
(\ref{canmomhampp6}) one obtains the following equations of motion 
for the probe brane:
\begin{eqnarray}
(\partial_t\rho)^2&=&{g_{tt}\over g_{\rho\rho}}\left[1-{{(P_{\phi}
-N\rho^{p+1})^2+T^2_pV^2_pe^{-2\Phi}(h_0r^2_0\rho^2)^pg_{\phi\phi}}
\over{E^2g_{\phi\phi}/g_{tt}}}\right],
\cr
(\partial_t\phi)^2&=&{{(P_{\phi}-N\rho^{p+1})^2g^2_{tt}}
\over{E^2g^2_{\phi\phi}}},
\label{r0eqmtn}
\end{eqnarray}
where $h_0\equiv h(r_0)$.  
The metric (\ref{idmetpm6}) on the brane universe is simplified to
\begin{equation}
d\tilde{s}^2=-{{T^2_pV^2_p}\over{E^2}}(h_0r^2_0\rho^2)^pe^{-2\Phi}
g^2_{tt}dt^2+h_0r^2_0\rho^2 d\Omega^2_p,
\label{bumet1}
\end{equation}
and therefore the cosmic time $\eta$ is defined through
\begin{equation}
d\eta^2={{T^2_pV^2_p}\over{E^2}}(h_0r^2_0\rho^2)^pe^{-2\Phi}
g^2_{tt}dt^2.
\label{cstmsph1}
\end{equation}
Substituting Eqs. (\ref{r0eqmtn}) and (\ref{cstmsph1}) into Eq. 
(\ref{lbzrslt}), one obtains the following form of the Friedman equation:
\begin{equation}
\left({\dot{a}\over a}\right)^2
={1\over \tilde{L}^2_p}\left[(\tilde{E}^2-\tilde{P}^2_{\phi})\left(
{\tilde{L}_p\over a}\right)^{2(p+1)}-\tilde{E}^2\left({\tilde{L}_p\over a}
\right)^{2p}+2\tilde{P}_{\phi}\left({\tilde{L}_p\over a}\right)^{p+1}-
\left({\tilde{L}_p\over a}\right)^2\right],
\label{frdeqsph1}
\end{equation}
where we introduced $(\tilde{L}_p/r_0)^4\equiv (L_p/r_0)^{p+1}$, $\tilde{E}
\equiv E\tilde{L}^2_p/(Nr_0)$ and $\tilde{P}_{\phi}\equiv P_{\phi}/N$ to 
simplify the expression.  This is the same form as the Friedman equation 
(\ref{frdeqsph}) describing the expanding spherical $p$-brane moving in the 
$S^{p+2}$ part of the AdS$_m\times S^{p+2}$ spacetime.  So, brane expansion  
in the near-horizon background of the dilatonic D-brane is qualitatively 
the same as that in the AdS$_m\times S^n$ background.  It is argued in 
Ref. \cite{dtv} that giant graviton can also be realized within backgrounds 
other than the AdS$\times S^n$ spacetime, provided some condition on the 
dilaton field and $h(r)$ in the metric in Eq. (\ref{nrdbrnmet}) is satisfied, 
as we summarize in the following. The Hamiltonian in Eq. (\ref{canmomhampp6}) 
can be expressed in terms of the conjugate momenta in the following form:
\begin{equation}
H=\sqrt{g_{tt}}\left[{{P^2_{\phi}}\over{h(r)r^2}}+
{{P^2_r}\over{g_{rr}}}+{{P^2_{\rho}}\over{g_{\rho\rho}}}+
{{(\rho P_{\phi}-N\rho^p)^2}\over{g_{\phi\phi}}}\right]^{1\over 2},
\label{harmcmdd}
\end{equation}
provided the condition $T_pV_pe^{-\Phi}(h(r)r^2)^{{p+1}\over 2}=N$ is 
satisfied.  This condition is satisfied by the solution (\ref{nrdbrnmet}) 
for the dilatonic D-brane in the near-horizon region.  For a given angular 
momentum $P_{\phi}$ and with $\partial_tr=0$, the Hamiltonian takes the 
following minimum value
\begin{equation}
H=\sqrt{g_{tt}}{{P_{\phi}}\over{r\sqrt{h}}},
\label{minhamdd}
\end{equation}
provided $\rho$ takes the equilibrium value (i.e. $P_{\rho}=0$) 
determined by
\begin{equation}
P_{\phi}=N\rho^{p-1}.
\label{rhophi}
\end{equation}
This minimum energy coincides with the energy of a massless particle, 
i.e. graviton, with angular momentum $P_{\phi}$ on $S^{p+2}$.  For our 
case, such minimum energy is given by
\begin{equation}
H=E={1\over r_0}\left({r_0\over L_p}\right)^{{p+1}\over 2}P_{\phi}=
{r_0\over\tilde{L}^2_p}P_{\phi},
\label{minendd}
\end{equation}
which coincides with the condition $|\tilde{E}|=|\tilde{P}_{\phi}|$ 
for the acausal term in the Friedmann equation (\ref{frdeqsph1}) to vanish, 
just as in the cases in the previous section.  

Second, we consider the $\partial_t\rho=0$ (i.e. $\rho=\rho_0={\rm const}$) 
case.  This case corresponds to the spherical probe $p$-brane with constant 
radius $\rho_0=(P_{\phi}/N)^{1/(p-1)}$ (i.e. the giant graviton) 
moving along the $r$-direction of the source brane.  The probe brane 
takes the following energy that coincides with the energy of a massless 
particle:
\begin{equation}
H=\sqrt{g_{tt}}\left[{P^2_{\phi}\over{h(r)r^2}}+{P^2_r\over{g_{rr}}}
\right]^{1\over 2}.
\label{mslspen}
\end{equation}
It is interesting to note that although the spherical brane has constant 
radius the brane universe evolves with time due to its motion along the 
$r$-direction in the source background (cf. the cosmic scale factor is 
given by $a=\sqrt{h(r)}r\rho_0$).  One can view the $\partial_t\rho=0$ case 
as a toy cosmological model on a fuzzy noncommutative sphere.  From Eq. 
(\ref{canmomhampp6}) with $\partial_t\rho=0$, one obtains the following 
equations of motion for the probe brane:
\begin{eqnarray}
(\partial_t r)^2&=&{g_{tt}\over g_{rr}}\left[1-{{(P_{\phi}-N\rho^{p+1}_0)^2+
T^2_pV^2_p(hr^2\rho^2_0)^pe^{-2\Phi}g_{\phi\phi}}\over{E^2g_{\phi\phi}/
g_{tt}}}\right],
\cr
(\partial_t\phi)^2&=&{{(P_{\phi}-N\rho^{p+1}_0)^2g^2_{tt}}\over
{E^2g^2_{\phi\phi}}}.
\label{r0eqmtn2}
\end{eqnarray}
The brane universe metric (\ref{idmetpm6}) is then simplified to
\begin{equation}
d\tilde{s}^2=-{{T^2_pV^2_p}\over E^2}(hr^2\rho^2_0)^pe^{-2\Phi}
g^2_{tt}dt^2+h^2r^2\rho^2_0d\Omega^2_p,
\label{bumet2}
\end{equation}
and therefore the cosmic time $\eta$ is defined through
\begin{equation}
d\eta^2={{T^2_pV^2_p}\over E^2}(hr^2\rho^2_0)^pe^{-2\Phi}
g^2_{tt}dt^2.
\label{cstmsph2}
\end{equation}
So, the Friedman equations (\ref{lbzrslt}) take the forms:
\begin{equation}
\left({\dot{a}\over a}\right)^2
={{(p-3)^2}\over{16L^2_p}}\left[\tilde{E}^2\tilde{P}^{-2{{p-1}\over
{p-3}}}_{\phi}\left({a\over L_p}\right)^{2{{p+1}\over{p-3}}}-\left({L_p\over 
a}\right)^2\right],
\label{frdeqsph2}
\end{equation}
\begin{equation}
{\ddot{a}\over a}={{(p-1)(p-3)}\over 8}\tilde{P}^{-2{{p-1}\over{p-3}}}_{\phi}
{\tilde{E}^2\over L^2_p}\left({a\over L_p}\right)^{2{{p+1}\over{p-3}}},
\label{2frdeqsph2}
\end{equation}
where we introduced $\tilde{E}\equiv EL_p/N$ and $\tilde{P}_{\phi}\equiv 
P_{\phi}/N$ to simplify the expressions.  
The effective energy density and pressure on the brane universe is 
therefore given by
\begin{eqnarray}
\varrho_{\rm eff}&=&{{p(p-1)}\over{256\pi L^2_p}}\left[(p-3)^2\tilde{E}^2
\tilde{P}^{-2{{p-1}\over{p-3}}}_{\phi}\left({a\over L_p}\right)^{2{{p+1}
\over{p-3}}}-(p+1)(p-7)\left({L_p\over a}\right)^2\right],
\cr
\wp_{\rm eff}&=&-{{p-1}\over{256\pi L^2_p}}\left[(p-3)(p^2-p+2)\tilde{E}^2
\tilde{P}^{-2{{p-1}\over{p-3}}}_{\phi}\left({a\over L_p}\right)^{2{{p+1}\over
{p-3}}}\right.
\cr
& &\ \ \ \ \ \ \ \ \ \ \ \ \ \ \ \ 
\left.-(p+1)(p-2)(p-7)\left({L_p\over a}\right)^2\right].
\label{effdensph2}
\end{eqnarray}
The scalar curvature (\ref{rscsph}) of the brane universe takes the 
following form:
\begin{equation}
{\cal R}={{p(p^2-1)}\over{16L^2_p}}\left[(p-3)\tilde{E}^2\tilde{P}^{-2{{p-1}
\over{p-3}}}_{\phi}\left({a\over L_p}\right)^{2{{p+1}\over{p-3}}}-(p-7)
\left({L_p\over a}\right)^2\right].
\label{crvsph2}
\end{equation}
From Eq. (\ref{frdeqsph2}), one can see that the brane universe cannot 
reach zero size (since the RHS becomes negative when $a=0$) when $p>3$.  So, 
the brane universe starts from some finite size where $\dot{a}=0$ and 
expands indefinitely (since $\ddot{a}$ is always positive).  [As in section 
3.2, despite having positive spatial curvature the brane universe expands 
indefinitely, because and $\varrho_{eff}$ and $\wp_{\rm eff}$ does not stay 
positive all the time during the expansion.]  
Also, what is extraordinary about this case is that the expansion of the 
brane universe accelerates with time, i.e., the brane universe inflates, 
rather than slowing down, since $\ddot{a}>0$ all the time.  Since the brane 
universe cannot reach zero size, the spacetime curvature (\ref{crvsph2}) 
remains finite at all time during the cosmological evolution.  Such violation 
of the singularity theorems of cosmology is possible because the mirage 
energy density and pressure do not satisfy the positivity conditions of the 
theorems.

\end{document}